\documentstyle[aps,prd,epsf]{revtex}
\begin{document}
\draft

\title{Center vortices, nexuses, and fractional topological charge}
\author{John M. Cornwall\thanks{Email address:  Cornwall@physics.ucla.edu}}
\address{ Department of Physics and Astronomy, University of California, Los Angeles, Los Angeles, California 90095}
\date{November 9, 1999}
\maketitle

\begin{abstract}

It has been remarked in several previous works that the combination of center vortices and nexuses (a nexus is a monopole-like soliton whose world line mediates certain allowed changes of field strengths on vortex surfaces) carry topological charge quantized in units of $1/N$ for gauge group $SU(N)$.  These fractional charges arise from the interpretation of the standard topological charge $\int G\tilde{G}$ as a sum of (integral) intersection numbers weighted by certain (fractional) traces.  We show that without nexuses the sum of intersection numbers give vanishing topological charge (since vortex surfaces are closed and compact).   With nexuses living as world lines on vortices, the contributions to the total intersection number are weighted by different trace factors, and yield a picture of the total topological charge as a {\em linking} of a closed nexus world line with a vortex surface; this linking gives rise to non-vanishing but {\em integral} total topological charge. This reflects the standard $2\pi$ periodicity of the theta angle.  We argue that the Witten-Veneziano $\eta^{\prime}$ relation, naively violating $2\pi$ periodicity, scales properly with $N$ at large $N$ without requiring $\theta$ periodicity of $2\pi N$.  This reflects the underlying composition of {\em localized} fractional topological charge, which are in general widely-separated.  Some simple models are given of this behavior.  In the intersection-number picture of topological charge, nexuses lead to non-standard surfaces for all $SU(N)$ and intersections of surfaces which do not constitute manifolds for $N>2$.  We generalize previously-introduced nexuses to all $SU(N)$ in terms of a set of fundamental nexuses, which can be distorted into a configuration resembling the 't Hooft-Polyakov monopole with no strings.  Nexuses can also be exhibited with thick non-singular strings, which generate vortices  with nexus (and anti-nexus) world lines appearing as boundaries on the vortex surface.  The existence of {\em localized} but widely-separated fractional topological charges, adding to integers only on long distance scales, has important implications for fermion zero modes and the existence of standard chiral condensates in chiral symmetry breakdown, avoiding the usual difficulty when the number of flavors exceeds one, as we review.


\end{abstract}

\section{Introduction}

A long-standing problem in gauge theories such as QCD is the existence or otherwise of fractional topological charge (quantized in units of $1/N$ for gauge group $SU(N)$).  On the one hand, the second Chern character ({\it i. e.}, the total topological charge) is integral for compact d=4 spaces.  On the other hand, not only has fractional topological charge been shown to exist in several circumstances (for the d=2 O(3) non-linear sigma model see Ref. \cite{ffs}; for torons, existing only in d=4 spaces not admitting fundamental fermions, see Ref. \cite{th81}; for $SU(2)$ charge-1/2 instantons joined by a sphaleron world line, see Ref. {\cite{ct}), but $1/N$ charges seem to be essential for the large-$N$ consistency of the Witten-Veneziano formula \cite{w79,v79} relating $\eta^{\prime}$ parameters to the topological susceptibility.  This consistency suggests that at large $N$ the partition function of the gauge theory, as a function of $\theta$, is:
\begin{equation}      
Z=\exp \int d^4x N^2F(\theta /N)
\end{equation} 
which in turn suggests a $\theta$ period of $2\pi N$.  (The extra $N^2$ in front comes from the known dependence of the vacuum energy at $\theta=0$.)

Another long-standing problem has to do with fermionic condensates, either
those related to chiral symmetry breakdown (CSB) or to gluino condensates in supersymmetric gauge theories.  If topological charge exists only in integral units, then the CSB condensate involves $N_f$ fermion bilinears, not what is observed for QCD, where there is a $\bar{q}q$ condensate.

In this work we argue that the combination of center vortices             \cite{c79,th79,mp79,no79} and nexuses\footnote{A nexus, reviewed in the next section, is the closest thing QCD has to a monopole.  It constitutes a closed world line sitting on a vortex, dividing the field strengths on the vortex into two or more distinct classes, as pointed out in Ref. \cite{ct}.  It is not the same as monopoles usually cited in gauge theories broken to $U(1)^{N-1}$.} \cite{c98,c99,agg} automatically solves the Witten-Veneziano problem in QCD, and also leads to an understanding of how chiral symmetry breakdown (CSB) works.

There is a closely-related problem in SUSY gauge theories, which we will not discuss at all in this paper beyond this paragraph.  However, it will be clear that the ingredients we use in the non-supersymmetric case are quite applicable to the supersymmetric case as well.  In the pure SUSY gauge theory with no fundamental-representation multiplet, the existence of fermionic zero modes is compatible with the existence of fractional total topological charge, because in the anomaly the $G\tilde{G}$ term is multiplied by an extra factor of $2N$ coming from the adjoint trace; the number of fermionic zero modes is even, by charge conjugation, so a total topological charge $\sim 1/N$ is compatible with an integral number of zero modes.  However, if it is assumed that only integral charge appears, there are $2N$ zero modes.  Nonetheless, one can argue (see Refs. \cite{s97,sm96} which include references to earlier works) that SUSY Ward identities and integral-charge instanton physics lead to a non-vanishing two-gluino (rather than $2N$-gluino) condensate $\langle \lambda \lambda \rangle$, with breakdown of a $Z_{2N}$ symmetry to $Z_2$, and the appearance of $N$ degenerate vacua, which are permuted with one another as $\theta$ changes.  These vacua are separated from each other by domain walls.  The alternative is a superselection rule,
according to which there is no possibility of connecting these degenerate vacua and no domain walls; which vacuum is to be used depends on the value of $\theta$
as it varies between 0 and $2\pi N$.  Some workers \cite{dhkm} claim that the assumption of clustering made in the SUSY-instanton calculations is not justified, and that the configurations contributing directly to  $\langle \lambda \lambda \rangle$ are of monopole type.  So there is controversy about the SUSY calculations, and it is not easy in broken SUSY to reveal what happens in conventional QCD, even if these controversies are resolved within SUSY.

Various  hypotheses have been made about the non-SUSY QCD case.  Witten \cite{w79} reconciles $2\pi$ periodicity with the $\eta^{\prime}$ physics by assuming the existence of $N$ non-degenerate vacua, one of which is picked out as having minimum energy at any given $\theta$; this is essentially the superselection rule option.  Others \cite{hz} believe the solution is rather like the SUSY one.

Our solution is similar in spirit to the assertions made \cite{dhkm,vb} for periodic instantons (calorons) that these can be decomposed into fractional topological charges such as monopoles, although in our case the fractional charges come from what appear to be different objects.  It remains to be seen whether there is any direct connection, when periodicity is not imposed and the gauge theory is not broken to $U(1)^{N-1}$.     

In view of these obscurities, one might ask whether there is any direct evidence from lattice calculations concerning the existence of fractional topological charge.  In fact, there is; Edwards {\it et al} \cite{ehn} find $SU(2)$ gauge configurations which have only two adjoint-fermion zero modes instead of the expected four, and interpret this as evidence of fractional topological charge. 
Exactly what carries this charge has  not yet been identified, except that it cannot be torons, which exist only in certain toroidal spaces not used by these authors.

Although center vortices have been around for twenty years, it is only recently that several groups \cite{tk,dfgo,fe} have found spectacular evidence that center vortices, and only center vortices, are the mechanism of confinement.
Moreover, de Forcrand and D'Elia give evidence that if center vortices are removed, CSB is lost; this is consistent with our arguments below about the role of center vortices and nexuses in CSB.

There is as yet only preliminary lattice evidence for the existence of nexuses; Ref. \cite{agg} presents both some arguments and some lattice data arguing for their existence in $SU(2)$.     

To continue elaborating on center vortices and nexuses as sources of fractional topological charge, we will give in further sections what we find is a  compelling picture of a vacuum structure in which it is automatic (at least for compact d=4 spaces) that the total topological charge is indeed integral, but it is composed of fractional sub-charges, quantized in units of $1/N$.  These fractional charges are separately localized (to within the characteristic distance $M^{-1}$ of QCD, where $M$ is identified with the thickness of center vortices) but are uncorrelated with one another except on a global distance scale, where the constraint of integral total topological charge operates.  As a result, the topological susceptibility, essentially the mean-square topological charge, is independent of $N$ at large $N$ (equivalent to the dependence of $F$ on $\theta/N$ in (1)) which insures the consistency of the Witten-Veneziano formula.  These results for the topological charge and its quantization stem from the observation \cite{c94,c96,c98,c99} that topological charge is essentially a sum of intersection numbers of the closed 2-surfaces which define center vortices.
These intersection numbers are weighted by traces coming from the flux matrices associated with center vortices and their co-existing nexuses, leading to fractional topological charge.    For simple topological reasons, the charge associated with the intersection of simple vortices having no nexuses is automatically zero, and essentially for the same reasons vortices with nexuses have integral and possibly non-zero topological charge.  In effect, the nexuses result in non-orientable surfaces\footnote{This effective non-orientability is not the same as actual non-orientability such as found in Klein bottles; there is lattice evidence \cite{bfgo} that center-vortex surfaces are highly non-orientable in the usual sense.} and in surfaces which are not manifolds.

Similar considerations hold for the Chern-Simons number in d=3, which is also \cite{c94,c96,c97} a linking number, in this case of d=1 closed loops, which is what vortices become in this dimension.  From time to time in the present paper we will use figures portraying, or speak of, d=3 vortices, but this is simply for ease of visualization; our interest here is in d=4.  We postpone to another work discussion of the d=3 gauge theory with Chern=Simons terms for several reasons; one is that the Chern-Simons term in the action affects the equations of motion, leading to twisted vortices \cite{cor96} carrying Chern-Simons number, a complexity with no analog in d=4.
Reconnection of vortices in d=3 may play an important role in B+L violation in the high-temperature early universe \cite{c97}.

Nexus world lines appear as boundaries on the closed vortex surfaces, dividing it into two or more parts.  In these different parts, the field strengths associated with center vortices differ, but they always give rise to the same vortex holonomy, expressed through the fundamental-representation Wilson loop as an element of the center $Z_N$ of the gauge group.  The intersection number of center vortices which expresses the topological charge is actually a (weighted) linking number of a vortex surface (with or without nexuses) with the world line of a nexus.  Of course, the fundamental property of confinement by center vortices is also expressed \cite{c79} through a linking number, in this case, that of the linking of the Wilson loop with the closed vortex 2-surface.
A vortex-loop link number can also be defined \cite{co96} for baryonic Wilson loops, which in $SU(N)$ have $N$ lines all of the same orientation going from one point to another, and which are not one-dimensional manifolds.  These baryonic linking numbers are rather similar in their general topological character to the linking numbers of certain vortices with nexuses.  

Another important characteristic of the topological charge carried by center vortices and nexuses has to do with CSB in QCD.  It is well-known that CSB with ordinary integrally-charged instantons leads to a condensate with $N_f$ fermion bilinears, where $N_f$ is the number of flavors.  This happens because of the $N_f$ fermionic zero modes, which are four-dimensionally renormalizable.  In fact, in order to get just a fermion bilinear condensate, it is necessary to have {\em three-dimensional} zero modes, that is, zero modes which extend indefinitely in a single direction in Euclidean four-space, but are localized in the other three directions \cite{gt86}.  Only such d=3 modes can lead to a non-zero and non-singular density of fermionic eigenvalues at the origin, which is necessary \cite{bc} for formation of a bilinear fermion condensate.  In effect, the fermions propagate along a line.  In the case of fractional topological charge, there is either explicitly (for $SU(2)$, see \cite{ct}) or implicitly such a line joining two or more fractional charges summing to an integer.  By the usual entropy arguments for strongly-coupled gauge theories such as QCD, this line is long and randomly-varying, and leads to no long-range correlations between the fractional charges except at the scale of the overall system size.  

There is one last question which should be discussed to complete the picture.  If conventional integrally-charged instantons coexist with the fractional charges discussed here, they will spoil the Witten-Veneziano formulas at large $N$.  In fact, there is no reason why this should happen.  In $SU(2)$ \cite{ct} the existence of 1/2-charge instantons (not the same as merons!) has been explicitly exhibited, along with the sphaleronic world line joining them.  Entropy considerations favor the decay of any unit-charge instanton into two 1/2-charge instantons.  Similar explicit pictures are not yet known for $N\geq 2$, but it is clear that once again a unit-charge instanton lowers the free energy by decaying into the kind of fractional-charge configurations we discuss here, which have high entropy.  There is no topological barrier to this kind of decay, which preserves the topological charge.  

In summary:  The fractional topological charges associated with center vortices and instantons are necessarily grouped into integral charges on a global scale, but are uncorrelated with each other on any shorter scale; the result is conventional $2\pi$ periodicity in $\theta$ along with Witten-Veneziano consistency.  Moreover, these fractional charges should lead to three-dimensional fermionic zero modes which in turn lead to the usual picture of CSB in QCD.

\section{Center vortices and nexuses}

We give a brief review of center vortices and nexuses, followed by a description for general $SU(N)$ of some fundamental nexuses from which all other nexuses can be found.  Examples of both fundamental ($SU(2)$) and composite ($SU(3)$) nexuses were given earlier \cite{c98}.  Each nexus, essentially a smeared-out point particle in d=3, or world line in d=4, must be accompanied by an anti-nexus to which it is joined by pieces of center vortices.  From time to time we will speak of a nexus in isolation, without reference to the anti-nexus to which it must be attached by vortex surfaces.

\subsection{  Review of center vortices and nexuses}

Center vortices and nexuses arise as quantum solitons of an effective action which reflects the main consequence of infrared instability of QCD in d=3,4, namely, generation \cite{c82,chk} of a constituent gluon mass $M$.  There is no symmetry breaking associated with the mass, which appears in the action as a gauged non-linear sigma model term.  Such a term, if interpreted as fundamental, can lead to non-renormalizability, but in fact no such problems are encountered because for consistency with Schwinger Dyson equations {\cite{c82} the constituent mass term in the gluonic propagator must vanish at large momentum \cite{ml}, in a way depending on the gluon condensate:
\begin{equation}     
M^2(q) \rightarrow const.\frac{\;g^2\langle G^2 \rangle}{q^2}
\end{equation}
modulo logarithmic corrections.  The massive effective Lagrangian density in Euclidean space is:\footnote{Until section IV we drop the coupling $g$; in fact, all vortex and nexus gauge potentials should have an extra factor $1/g$.  The gauge potential is the usual anti-Hermitean one:  $A_{\mu}=(\lambda_a/2i)A_{\mu}^a$ with Tr$\lambda_a\lambda_b=2\delta_{ab}$, and the covariant derivative is $D_{\mu}=\partial_{\mu}+A_{\mu}$.}
\begin{equation}       
{\cal{L}}=(-Tr)[\frac{1}{2}G_{\mu \nu}^2+M^2(U^{-1}D_{\mu}U)^2].
\end{equation}
Here $U$ is an $N\times N$ unitary matrix; the gauge transformation laws
\begin{equation}      
U\rightarrow VU;\;A_{\mu}\rightarrow VA_{\mu}V^{-1}+V\partial_{\mu}V^{-1}
\end{equation}
leave both terms of the Lagrangian invariant.   It can easily be checked that the equations of motion for $U$ are not independent, but merely represent the usual identity 
\begin{equation}     
[D_{\mu},[D_{\nu},G_{\mu\nu}]]\equiv 0.
\end{equation}

Both center vortices and nexuses are quantum solitons of the effective action (3).  We will postpone the discussion of nexuses to the next subsection, for now noting that nexuses are monopole-like objects on which two or more vortex sheets terminate.

A center vortex with no nexus structure on it can always be displayed (modulo a regular gauge transformation) as an Abelian structure, representing the underlyling Abelian (center) holonomy of the vortices.  (Once nexuses enter the picture, non-Abelian effects must appear.)  The essential contribution of the $U$ terms is a long-range pure-gauge part which is in itself singular on a closed 2-surface in d=4 (a closed string in d=3).  In d=4, a solitonic solution \cite{c79} to the effective action based on equation (3) is essentially a Nielsen-Olesen vortex:
\begin{equation}      
A_{\mu}(x;Q)=(\frac{2\pi Q}{i})\epsilon_{\mu\nu\alpha\beta}\partial_{\nu}\int \frac{1}{2}d\sigma_{\alpha\beta} [\Delta_M(x-z)-\Delta_0(x-z)].
\end{equation}
Here $Q$ is a matrix in the Cartan subalgebra (discussed below), the integral runs over a closed 2-surface $z(\sigma ,\tau)$, and $\Delta_{M,0}$ is the free propagator of mass $M,0$ respectively.  The $U$ contribution is the $\Delta_0$ term, which can easily be verified to be a pure gauge.  By itself this term is singular when $x=z$, but this singularity is exactly cancelled by the massive term.  

In this paper we will only be interested in long-range effects, so for the most part we take $M\rightarrow \infty$, drop the $\Delta_M$ term, and save only the long-range pure-gauge part.  When this is done the resulting field strength is not identically zero, but has delta-function strength on the vortex surface.  (In this limit these become Dirac strings for nexuses.)   It is simplest to express this through the dual field strength:
\begin{equation}     
\tilde{G}_{\mu\nu}=(\frac{2\pi Q}{i})\int d\sigma_{\mu\nu}\delta (x-z).
\end{equation}     
 It is convenient in the $M=\infty$ limit to observe \cite{c98} that this pure-gauge part can be written as:
\begin{equation}      
A_{\mu}(x;Q)=V\partial_{\mu}V^{-1};\;V=\exp i\Omega Q
\end{equation}               
where $\Omega$ is the so-called magnetic potential, defined as an integral over any 3-volume $\cal{V}$ whose boundary $\partial \cal{V}$ is the surface $S$ over which the vortex is defined:
\begin{equation}      
\Omega (x) =  \int dV_{\mu}\partial_{\mu}\Delta_0(x-z).
\end{equation}
The delta-function field strength is reproduced via the $2\pi$ jump in $\Omega$ as the volume $\cal{V}$ is crossed.

The matrix $Q$ which appears in the vortex (6) we will call a {\em flux matrix}. Flux is defined not through a $\Pi_2$ homotopy, as is often asserted for the Abelian monopole case, but by the homotopy $\Pi_1(SU(N)/Z_N)=Z_N$.  In physical terms, the magnetic flux is measured by a Wilson-loop holonomy going around a vortex string.
For gauge group $SU(N)$ the matrix $Q$ is chosen as a sum of the $N$ matrices
$Q_i$, which (over)span the Cartan subalgebra (the sum of all $Q_i$ is zero):
\begin{equation}     
Q_i=\;diag(\frac{1}{N},\;\frac{1}{N},\;\frac{1}{N}-1,\;\frac{1}{N},\dots )
\end{equation}       
where the -1 is in the $i^{th}$ position.  Any of the matrices $Q_j$ has the property that $\exp (2\pi iQ_j)=\exp (2\pi i/N)$, an element of the center $Z_N$.  The sum $Q_i+Q_j+\dots$ with $i\neq j\neq \dots$ and a total of $k$ terms exponentiates to the center element $\exp (2\pi ik/N)$.

To describe the effect of vortices on (fundamental) Wilson loops, it is not necessary to be so explicit.  A set of independent vortices is described by an overall gauge transformation
\begin{equation}      
V=V(k_1)V(k_2)\dots V(k_J)
\end{equation}
where the $V(k_j)$ have holonomies $\exp (2\pi ik_j/N)$, when linked once in a positive sense with a given Wilson loop.  For a given set of vortices, the value of the Wilson loop is given by the product of these holonomies, each counted as many times as the vortex is linked with the loop.  (If the Wilson loop is unlinked to a given vortex, that vortex contributes a factor of unity.)  A standard collective-coordinate calculation for a gas of independent vortices then gives \cite{c79,co98} for the expectation value of the Wilson loop:
\begin{equation}     
\langle Tr\;P\exp [i\int dx_{\mu}A_{\mu}(x)]\rangle =\exp [\frac{\rho A}{N-1}\sum (e^{2\pi ik/N}-1)]
\end{equation}
where $A$ is the minimal area of a surface spanning the loop, $\rho$ is the two-dimensional density of vortices of all types penetrating unit area, and the sum over $k$ goes from $-[N/2]$ to $+[N/2]$, but excluding zero, where the brackets indicate the integral part.  One then finds a string tension $K_F=\rho N/(N-1)$. 

Note that the result for the value of the Wilson loop in a specified set of vortices (as in equation (11)) is just the same as if we had actually used the explicit Abelian form of the vortices as given in equation (6).  One sees here an application of the splitting principle \cite{egh} according to which gauge-invariant functions such as Wilson loops and characteristic classes can be computed as if the gauge connections were diagonal (living in the Cartan subalgebra $U(1)^{N-1}$ instead of in $SU(N)$.)  Later we will use this Abelian form to calculate topological charge from the Chern second character.  In this application nexuses appear as obstructions to global applicability of the splitting principle.

\subsection{  Fundamental vortex-nexus for $SU(N)$  }

A nexus is a physical branch point at which up to $N$ vortex surfaces can join.  Because these surfaces must be closed, a nexus can only exist along with an anti-nexus, as shown for $SU(2)$ in  Fig.\ \ref{fig1}  and for $SU(3)$ in Fig.\ \ref{fig2} ; the nexus or anti-nexus is shown as a dark circle where vortices meet.  These particular examples were discussed in detail in Ref. \cite{c98}.  The general rule for choosing the flux matrices on the individual vortices of the nexus is that the algebraic sum of all the incoming flux matrices must have only integral eigenvalues.  As we will show here, the simple loop topology of the $SU(2)$ nexus is the generic type for all $N$, and other vortex-nexuses, such as in Fig.\ \ref{fig2}, can considered to be composites of the fundamental type of Fig.\ \ref{fig1}, with two vortices, a nexus, and an anti-nexus. There is also what appears to be another type of vortex-nexus combination built on a simple closed loop;  this generalization for $N>2$ consists of a  simple closed loop, as in Fig.\ \ref{fig1}, with vortices divided by up to $N$ nexuses.  Each  vortex segment carries a different flux matrix $Q_i$.  We leave it to the reader to show that this is a composite of fundamental vortex-nexuses. 

A vortex-nexus combination which is a composite of fundamental ones may have different effective action from that ascribed to the fundamentals of which it is a composite; we do not discuss this purely dynamical point here.

Fig.\ \ref{fig1} and Fig.\ \ref{fig2} can be interpreted as composed either of one-dimensional strings, as appropriate for d=3 vortices, or as cross-sections of 2-surfaces, as appropriate for d=4.   In the next section we discuss the d=4 surfaces more explicitly, but
for simplicity of visualization think of d=3 nexuses in what follows.  It is most important to note that at finite $M$ these strings are {\em not} Dirac strings; rather, they are fat strings, in fact identical in local structure to the center vortices already introduced in equation (6).  Roughly speaking (for more detail, see \cite{c98}) the arrows on the vortex lines indicate the direction of the magnetic field of the vortex, and in Fig.\ \ref{fig2} the numbers indicate which $Q_i$ is carried on the labeled line.  The sum of the incoming flux matrices at the nexus of Fig.\ \ref{fig2} is therefore $Q_1+Q_2+Q_3=0$.  For the $SU(2)$ case of Fig.\ \ref{fig1}, no labels are needed, because the only possibility is that one line carries $Q_1=(1/2)\tau_3$, and the other carries $Q_2=-Q_1$.  In this case the sum of incoming fluxes is $Q_1-Q_2=2\tau_3$.  This is an allowed flux matrix at the nexus, because this flux matrix has only integral eigenvalues.  

As stated above, the general $SU(N)$ nexus obeys the simple rule that the sum of incoming flux matrices at any nexus must yield a flux matrix with only integral eigenvalues.  So Fig.\ \ref{fig2} could be relabeled; for example, all lines could carry $Q_1$ since
$3Q_1$ is integral.  It is only such nexuses, carrying {\em non-vanishing} but integral flux which can contribute to non-zero topological charge.  Note that however we choose the $Q_i$ on the vortex lines, the holonomies on all lines are the same.  Of course, more complicated nexuses can be constructed, in which every line has a flux matrix of the type $Q_i+Q_j$, with different values of $i,j$ for each line; again, the holonomies are the same for every line.  

Every $SU(N)$ nexus can be written in terms of a set of fundamental nexuses which have just two lines, as in the $SU(2)$ case of Fig.\ \ref{fig1}.  On one of the lines the flux matrix is, say, $Q_i$; on the other line the flux matrix is
$-Q_j$.  The only non-trivial case is $i\neq j$.   We will use the notation $(i,-j)$ for a fundamental nexus, of which there are clearly $N(N-1)/2$ different varieties.  The total flux matrix is
$Q_i-Q_j$, which is easily checked from equation (10) to have all but two eigenvalues of zero, the other two being 1 and -1.  That is, {\em for every $N$ and every $i\neq j$ the matrix $Q_i-Q_j$ is twice a generator $J_3$ of an embedded $SU(2)$ subgroup.}\footnote{By incorporating a zero eigenvalue of $Q_i-Q_j$ one has a $J_3$ for an embedded $SO(3)$; this works just as well.}  This is a key observation, because it allows us to deform a fundamental nexus for any $SU(N)$ into something resembling a 't Hooft-Polyakov (TP) monopole (except that there are no adjoint Higgs fields to break the symmetry to $U(1)^{N-1}$).  Alternatively, of course, we may think of deforming the TP-monopole of Ref. \cite{c99}, embedded in any $SU(N)$ gauge theory, into a fundamental nexus (and then closing all the vortex lines involved with an anti-nexus).

Consider now an $SU(3)$ three-line nexus, such as shown in Fig.\ \ref{fig2}; in an obvious notation, we label such nexuses generically $(i,j,k)$, so Fig.\ \ref{fig2} is the nexus $(1,2,3)$.  By drawing pictures and using the fact that the sum of the $Q_i$ is zero, one sees that it can be written as a composite of two fundamental nexuses (along with their partner anti-nexuses) of type $(1,-2)$, such that two of the lines are merged (beginning at the merger of the nexuses) in such a sense that the merged lines carry flux matrix $Q_3$.
Similarly one can compound all allowed nexuses out of fundamental ones.          

We return to the mapping of a nexus with integral flux onto the TP monopole.  This has been discussed in detail in Refs. \cite{c98,c99} for the $SU(2)$ case, so we can be brief about the generalization to $SU(N)$.  Consider the $(-1,2)$ nexus of $SU(3)$, shown in isolation from the anti-nexus in Fig.\ \ref{fig3}(a).  The vortex lines in this figure run along the $z$-axis.

 The first step in constructing this nexus is to determine the long-range pure-gauge part of the vector potential, which must be, in the notation of the effective Lagrangian of equation (3), $U\partial_{\mu}U^{-1}$ in order that both terms of the effective action go to zero at long distances.  In the $M\rightarrow\infty$ limit this is all that is left of the gauge potential.  The necessary gauge part must describe a $Q_1$ string on one side and a $Q_2$ on the other.  One simple possibility, by no means the only one, is:
\begin{equation}     
U=U_1=e^{i(\theta/2)\lambda_3}e^{i\phi Q_1}e^{-i(\theta/2)\lambda_3}.
\end{equation}
Here $\phi$ is the usual azimuthal angle and $\theta$ the usual polar angle.
It is easy to verify that this $U_1$ leads to Dirac strings as shown in Fig.\ \ref{fig3}(a).

The next step is to transform $U_1$ into a nexus such as is shown in Fig.\ \ref{fig3}(b), where the (-1) string has been moved onto the same side of the $z$-axis as the (2) string.  This is easily done by the transform
\begin{equation}     
U_1\rightarrow U_2=e^{-i\phi Q_1}U_1e^{i\phi Q_1}.
\end{equation}

The result is a monopole-like object with a half-string whose flux matrix is twice a $J_3$ generator of an embedded $SU(2)$, as described above; for $SU(3)$, this generator is $(1/2)\lambda_3$.  This will be recognized as the presentation of the Wu-Yang monopole in the so-called Abelian gauge.  At this point we will not bother to write out terms in the gauge potential which do not have Dirac strings.  If we take the half-string to lie along the negative $z$-axis, the corresponding vector potential describing the strings is:
\begin{equation}    
A_4=0;\;A_i=\frac{i}{\rho}\lambda_3\hat{\phi}(1-\cos \theta ).
\end{equation}
Of course, the double flux comes from the $1-\cos \theta$ factor along the $z$-axis.  
By a well-known gauge transformation this can be transformed into a spherically-symmetric gauge potential with no string:
\begin{equation}    
U_2\rightarrow VU_2V^{-1};\;V=e^{-i\theta \vec{J}\cdot\hat{\phi}}.
\end{equation}
Here $\vec{J}$ are the usual $SU(2)$ generators $(1/2)(\lambda_1,\lambda_2,\lambda_3)$ embedded in $SU(3)$.
The result is the familiar presentation of the Wu-Yang monopole in the spherical gauge, with no strings:
\begin{equation}      
A_4=0;\;A_i=\frac{-1}{2i}\epsilon_{ijk}\hat{r}_j\lambda_k.
\end{equation}
This form of the Wu-Yang potential is, in fact, the asymptotic behavior of the TP monopole in the spherical gauge.

We can now refer to Ref. \cite{c98} for further developments, in which the Wu-Yang potential is augmented to account for finite=$M$ terms needed to remove short-distance singularities and turn the original Dirac strings into fat strings without singularities, such as occur in center vortices.  Ref. \cite{c99} gives the explicit results needed to deform the TP monopole, originally defined through a theory containing adjoint Higgs fields, into one which has no Higgs fields but only the symmetric mass term of equation (3).  The resulting monopole has no long-range fields and no strings.
By  inverting the series of gauge transformations given above (suitably deformed to avoid singularities), we go from the spherically-symmetric monopole to the nexus of Fig.\ \ref{fig3}(a).  That is, first one puts in the half-string by means of the gauge transform in (16), then converts this into a full string as in Fig.\ \ref{fig3}(a) with the gauge transform of (14).  The result is not exactly the original gauge we started with, as shown in equation (13), but is equivalent to it.  When these steps are taken in the $SU(2)$ case the resulting nexus gauge can be chosen as:
\begin{equation}    
V=\exp (i\frac{\phi}{2}\vec{\tau}\cdot\hat{r});\;A_i=V\partial_iV^{-1}.
\end{equation}
One easily finds that the singular field strength points along the $z$-axis, but with oppositely-directed field strengths, as shown in Fig.\ \ref{fig1}.  This happens because $\vec{\tau}\cdot\hat{r}=\pm\tau_3$ along the $z$-axis.  One also sees that, as discussed above, the holonomy around either of the two strings attached to the nexus is -1. 
 
The static $SU(3)$ nexus of Fig.\ \ref{fig2}, with three strings, can be described analogously (see Ref. \cite{c98}), but we need not review that here.   It is enough for us to know that this vortex is a composite of two $(i,j)$ vortices, as described above. 

\section{Topological charge}

We first show that compact closed vortices with no nexuses give rise to zero topological charge, and next that adding nexuses gives rise to non-vanishing integral topological charge.  In both cases the essential feature is that the topological charge integral $\int G\tilde{G}$ is expressible as a (weighted) intersection number.

\subsection{Topological charge as an intersection number}

First we define an integral which gives intersection numbers, each weighted by an orientation factor.  Let $S_{A,B}$ be two closed, oriented, compact 2-surfaces in d=4 and define an intersection integral $I(A.B)$ by:
\begin{equation}     
I(A,B)=\epsilon_{\mu\nu\alpha\beta}\int \frac{1}{2}d\sigma_{\mu\nu}(S_A)\frac{1}{2}d\sigma_{\alpha\beta}(S_B)\delta (x(A)-x(B)).
\end{equation}
 
The topological charge $Q$ is:
\begin{equation}     
Q=\frac{-1}{16\pi^2}\int d^4x Tr\;G_{\mu\nu}\tilde{G}_{\mu\nu}.
\end{equation} 
With the help of equations (19) and (7) (giving the dual field strength of a center vortex), this results in a formula for the topological charge coming from a number of oriented vortices (for the moment, with no nexuses): 
\begin{equation}    
-Q=\sum_{A,B} Tr (Q_AQ_B)N(A,B).
\end{equation}
Here $N(A,B)$ is the number of times, weighted by orientation, that surface $A$ intersects with surface $B$.  This number is actually twice the value $I(A,B)$ of the integral in (19).  The reason is that when one expresses the topological charge as a sum over a quadratic form in vortex surfaces, mutual intersections are counted twice.  In fact, so are self-intersections, which are counted twice in $I(A,A)$.  The reason is that on the surface $A$ we can identify two neighborhoods, call them 1 and 2, which contain a point of intersection.  Just like mutual intersections, these bits of surface are counted twice in $I(A,A)$.

It would now appear from (21) that the {\em total} topological charge could be non-integral, because of the trace factors.  This is not so, for simple vortices with no nexuses.  The reason is that the intersection number for each trace is actually zero, reflecting the fact that closed surfaces intersect an even number of times, and the orientation factor is of one sign for half the intersections and the other for the other half (see Fig.\ \ref{fig4}, which shows the same thing but in d=2).

The conclusion is that for compact closed vortex surfaces with no nexuses, the topological charge is zero.  We next consider the influence of nexuses.

\subsection{Vortices, nexuses, and integral topological charge}

Nexuses modify equation (7), which expresses the field strength of a vortex with no nexuses as $Q$ times an integral over a delta function.  In the $M=\infty$ limit, the necessary modification to describe a vortex-nexus combination is that the matrix $Q$ becomes a matrix-valued function $Q(\sigma ,\tau)$ of the variables defining the surface $x_{\mu}=z_{\mu}(\sigma ,\tau)$:
\begin{equation}     
\tilde{G}_{\mu\nu}=(\frac{2\pi}{i})\int d\sigma_{\mu\nu}Q(\sigma ,\tau)\delta (x-z).
\end{equation}
This function $Q(\sigma ,\tau)$ changes on the scale $M^{-1}$ from $Q_i$ to $Q_j$  as one crosses the world-line of a fundamental nexus, which is embedded in the vortex surface.\footnote{As we will see below, one cannot choose the $\sigma ,\tau$ dependence of $Q$ arbitrarily; the Bianchi identities must be respected.}  For example, consider the d=3 nexus of Fig.\ \ref{fig1}, applicable to $SU(2)$; the direction of field strengths changes at the nexus or anti-nexus.  To make a closed vortex surface in d=4, rotate the nexus and the anti-nexus to form two closed world lines, which appear (see Fig.\ \ref{fig5}) as two boundaries dividing a torus into two parts, with oppositely-directed field lines in each part.  Although not shown on the figure, the nexus world lines are oriented, and oriented oppositely to each other.

If the vortex-nexus combination in Fig.\ \ref{fig4} intersects a plain vortex with no nexuses, it is possible to generate non-zero topological charge, provided that one of the two intersection points lies in one region of the vortex-nexus, and the other lies in the other region, where the field lines have opposite orientation.  This is illustrated in Fig.\ \ref{fig6}, where a slice of the plain vortex is shown intersecting the vortex-nexus combination.

When there were no nexuses, as in equation (21), we encountered the trace weighting factor
\begin{equation}     
Tr(\frac{\tau_3}{2i})^2-Tr(\frac{\tau_3}{2i})^2=0.
\end{equation}
The minus sign reflected the fact that the vortex slice of Fig.\ \ref{fig5} first entered, then exited the torus.
But now, provided that the two intersection points lie in the two different regions of the torus, we find
\begin{equation}      
Tr(\frac{\tau_3}{2i})^2-Tr[-(\frac{\tau_3}{2i})^2]=1.
\end{equation}
The minus sign inside the trace comes from the reversal of field strengths.

It is clear that the topological charge is now expressed as a {\em linking number of a vortex with the world line of a nexus}.  When this linking number is zero, the topological charge is zero; this can happen either if there is no nexus (as in the previous subsection} or if the two intersection points lie on the same side of the nexus, that is, are not linked to it.  

All of this generalizes to $SU(N)$.  In that case, if a plain vortex $(k$) is linked to the nexus $(i,-j)$ of another vortex, the trace factor which we encounter is:
\begin{equation}       
TrQ_k(Q_i-Q_j)=\delta_{ik}-\delta_{jk}.
\end{equation}
The result is always $\pm 1$ or 0.  The reader can supply the generalization to a vortex-nexus combination intersecting another vortex-nexus combination, to higher linking numbers, and so on.  In all cases, the topological charge is an integer.  

Non-fundamental nexuses involve higher-genus figures which are awkward to draw in d=3 (just as it is awkward to draw a Klein bottle).  For example, an $(i,j,k)$ vortex-nexus, as in Fig.\ \ref{fig2}, requires a genus-two surface which is not a manifold.  

We close this section with some remarks on the formula (22), which generalizes to a nexus-vortex combination the fundamental $M=\infty$ vortex of equation (7).
With $Q$ a non-trivial function of $\sigma ,\tau$, the Bianchi identity is not automatic.  To simplify the issue, we note that $Q$ changes only in the immediate vicinity of the nexus line, which can be parametrized with only one of the two variables $\sigma ,\tau$.  For example, choose these 2-surface coordinates so that the nexus world line is $\tau =0$; that allows us to choose $Q$ to be only a function of $\tau$.  Applying the Bianchi identity one finds, using
\begin{equation}     
d\sigma_{\mu\nu}=\dot{z}_{\mu}z^{\prime}_{\nu}-(\mu \leftrightarrow \nu),
\end{equation}
where the dot indicates $\tau$ derivatives and the prime indicates $\sigma$ derivatives, that
\begin{equation}     
\partial_{\mu}\tilde{G}_{\mu\nu}=(\frac{2\pi}{i})\int dz^{\prime}_{\nu}\{
\dot{Q}+[\dot{z}_{\mu}A_{\mu},Q]\}\delta (x-z).
\end{equation}
In order that the Bianchi identity be satisfied, the quantity in curly brackets on the right-hand side of (27) must vanish (Wong equation), so that
\begin{equation}     
Q(\tau )=U(\tau )Q(0)U^{-1}(\tau );\;U(\tau )=P\exp (-\int^{\tau}\dot{z}_{\mu}A_{\mu}).
\end{equation}
One can then write the vortex-nexus combination as a singular gauge transformation of the pure vortex:
\begin{equation}      
G_{\mu\nu}\rightarrow U(x)G_{\mu\nu}U^{-1}(x)
\end{equation}
where $U(x)$ is some suitable extension of $U(\tau )$ from a function defined on the vortex surface to one defined everywhere.\footnote{For example, the surface $x=z(\sigma ,\tau)$ can be described as the simultaneous satisfaction of two conditions $G(x)=H(x)=0$, for suitable $G,H$.  Then one can choose $G,H,\sigma ,\tau$ as four coordinates, expressible in terms of the $x_{\mu}$.  Any function $U(G,H,\sigma ,\tau )$ such that $U(0,0,0,\tau )=U(\tau )$ is an extension to all space.}  Such a singular gauge transformation is of the type described in equation (13), which rotates $Q_1$ into $Q_2$.  This transformation necessarily does not commute with either of the $Q$-values on the two parts of the vortex-nexus surface.

\section{Some simple models}

So far we have established that total topological charge is integral, but its constituent charges are fractional (which reminds one of conventional quark confinement).  This is the clue to devising models of the dependence of the vacuum energy on $\theta$ which are consistent with the Witten-Veneziano formula (1).

Let us recall equation (1) (repeated for convenience) which expresses the partition function in terms of an effective action:
\begin{equation}       
Z=\exp \int d^4x N^2F(\theta /N)\equiv \exp [-\int d^4x W(\theta )].
\end{equation}
In the first model, which is purely mathematical, we seek a function $F$ which is both (formally) periodic with period $2\pi$ and which has a second derivative at the origin which is $O(1/N^2)$ for the specific case $N=2$.  The second model is quasi-physical; for it, we estimate the large-$N$ behavior of the first two terms in the Taylor series expansion of $W$.  Our estimates are not quantitatively accurate, but we believe that they give the right large-$N$ behavior.

 \subsection{A mathematical model mimicking SU(2)}

Suppose we flip a coin at random, assigning the number +1/2 to every head and -1/2 to every tail.  There is a total of $J=N_++N_-$ trials, with $N_+$ heads and $N_-$ tails.  The ``topological charge" $Q$ is defined as:
\begin{equation}     
Q=\frac{1}{2}(N_+-N_-)=N_+-\frac{1}{2}J.
\end{equation}
Evidently if we choose $J$ even we always get integral topological charge, so set $J=2K$.  

It is elementary to calculate the generating function $Z(\theta )$ for this distribution:
\begin{equation}     
Z(\theta )=\sum \frac{J!}{N_+!(J-N_+)!}(\frac{1}{2})^Ne^{i\theta (N_+-K)}=(\cos \frac{\theta}{2})^J.
\end{equation}
Since $J$ is even we can also write 
\begin{equation}     
Z(\theta )=(\frac{1+\cos \theta )}{2})^K,
\end{equation}
which shows that $Z$ is periodic in $\theta$ with period $2\pi$.

Write $Z(\theta )$ in terms of an ``effective action density" for topological charge, without the factor $N^2=4$ in front which we have agreed to drop:
\begin{equation}     
Z(\theta )=e^{-JF(\theta )};\;F(\theta )=(\frac{-1}{2})\ln (\frac{1+\cos \theta }{2}).
\end{equation}
Here $J$ plays the role of the system space-time volume (see equation (1)).  
We calculate the mean-square topological charge, corresponding to $1/N^2$ times the topological susceptibility in the Witten-Veneziano formula:
\begin{equation}     
\frac{\langle Q^2\rangle}{J}=-F^{\prime\prime}(\theta )|_{\theta =0}=\frac{1}{4}.
\end{equation}
The 1/4 is, of course, the factor $1/N^2$ we want to have in the Witten-Veneziano formula (1), coming from the second derivative of $F(\theta /N)$ introduced in that formula.  So we have a model in which the effective potential density is formally a periodic function of $\theta$ with period $2\pi$, while at the same time the susceptibility reflects the underlying compositon of the topological charge as a sum of fractional charges.

There is, as one might anticipate, a singularity in the effective potential density at $\theta=\pi$, where $F$  is logarithmically singular and goes to    
$+\infty$.  There is an infinite barrier to cross at $\theta =\pi$.  Once the barrier is crossed, the effective potential is well-defined in the region $\pi < \theta \leq 2\pi$.

Similar models can be constructed for larger $N$, but we will not discuss them here.   In every case, the trick is the occurrence of a logarithm in the effective action, leading to impassable barriers at specific values of $\theta$.

Let us now turn to  more physical arguments.

\subsection{A nexus-vortex picture}

The object here is to argue for a picture of large-$N$ nexus-vortex dynamics leading to consistency between the behavior of the vacuum effective-action density at $\theta=0$, which scales (as equation (1) or (30) shows) with $N^2$, and its second derivative, which should scale as $N^0$ to agree with the Witten-Veneziano formula.  In achieving this consistency we require that globally only integral topological charge is allowed.  That is,
the dynamics should be consistent with $2\pi$ periodicity of $\theta$.  The dynamics of nexuses and vortices in d=4 is very complicated, and all of our considerations will be qualitative and heuristic.   We do not (as we did in the previous model) determine the full $\theta$ dependence of the effective action, which is a difficult problem even in a drastically-simplified model.

The picture we offer is supposed to summarize in a heuristic way the results of some elaborate calculation of the vortex-nexus vacuum, including proper treatment of collective coordinates, soliton-soliton interactions, and the like.
As part of this calculation one would determine a density $\rho$ of solitons per unit 4-volume.  Schematically, this density is determined by:
\begin{equation}     
\rho=\sum_{c.c.} e^{-I}
\end{equation}
where the sum is over collective coordinates for the solitons, whose action is $I$.  Both this collective-coordinate sum and the action depend on $N$ and on the specific type of soliton.  For example, a vortex whose holonomy is $\exp (2\pi iJ/N)$ has an action (per unit vortex area) which depends on $N,J$ as:
\begin{equation}     
I/A\simeq (\frac{M^2}{g^2})(\frac{J(N-J)}{N})
\end{equation}
(where the factor $J(N-J)/N$ is $TrQ^2$ for the appropriate sum over different $Q_i$)
and so it would appear that the density $\rho$ goes to zero in the large-$N$ limit.  But this need not be so (see, {\it e.g.}, \cite{co98}) because the collective-coordinate integrals over group cosets contains factors which grow exactly at the rate at which $\exp (-I)$ decreases.  For extended objects like vortices it is beyond our powers to do the true collective-coordinate integrals, but it is plausible \cite{co98} that the density $\rho$ scales like a power of $N$ and is not exponentially-small.  In any event, that is what we assume.
We will see that for certain fundamental vortices, which is all that we consider explicitly, the scaling $\rho \sim N$ is needed to reproduce the $N^2$ scaling of the vacuum energy at $\theta=0$ (see equation (1) or (30)).  We then show that this scaling leads to a topological susceptibility which scales like $N^0$.

There are many types of vortex-nexus combinations which can inhabit the vacuum, and to discuss them all explicitly is more tedious than instructive.  We will only consider fundamental vortices with just one nexus-anti-nexus pair on it, as  illustrated in Fig.\ \ref{fig5}; all other types are composites of this case.  Then we can restrict the sum over vortices, collective coordinates, and so forth in equation (37) just to the case $J=1$, the case of fundamental vortex-nexus combinations.  From equation (37) the action of a fundamental vortex scales like $N$, and the effective-action density of the vacuum will scale like $\rho N$ after doing the collective-coordinate calculations.  This must be an $N^2$ scaling, which suggests, as already mentioned, $\rho \sim N$.

Previous sections of this paper have shown that the total topological charge $Q$ is necessarily integral for vortex-nexus combinations, and no approximation we have made will change that.  The question is what happens to the topological susceptibility $\langle Q^2\rangle /V$ (where $V$ is the volume of space-time).

By standard arguments of random sign and independence of the fractional topological charges $q_A$ which live on the intersection points $A$ of vortex-nexus combinations, the susceptibility depends only on the sum of squares of these fractional charges:
\begin{equation}     
\langle [\sum q_A]^2\rangle =\langle \sum (q_A)^2\rangle.
\end{equation}
Each $q_A^2$ in this sum is of the form $TrQ_iQ_j$, for various values of $i,j$.  Evidently $i=j$ with probability $1/N$ and $i\neq j$ with probability $(N-1)/N$.  In view of the trace formula, valid for all $N$,
\begin{equation}     
TrQ_iQ_j = \delta_{ij} -\frac{1}{N}
\end{equation} 
it follows that the mean-square topological charge per intersection of the fundamental vortex-nexus combinations is:
\begin{equation}  
\langle q_A^2\rangle = \frac{1}{N}(\frac{N-1}{N})^2+(\frac{N-1}{N})\frac{1}{N^2}= \frac{N-1}{N^2}.
\end{equation}
The topological susceptibility is found by multiplying by the density of intersections, which we argue scales like the soliton density $\rho$:
\begin{equation}    
\frac{\langle Q^2\rangle}{V}\sim \rho \langle q_A^2\rangle \sim N^0.
\end{equation}
This is the desired result for consistency with the Witten-Veneziano formula.

\section{Conclusions}

The thrust of this paper is that a theory like QCD does possess fractional topological charge, quantized in units of $1/N$ for gauge group $SU(N)$.  Becauses of the special topology of vortices with nexus world lines embedded in them, the interpretation of total topological charge as an intersection number does not always yield zero, as it would for pure vortices with no nexuses, but instead yields integral topological charge.  At the same time, the fractional topological charges are essential for consistency with the Witten-Veneziano formula.  It is not that (in general) all fractional charges are $1/N$, which would require $N-$fold intersections of vortices not describable as manifolds.
Instead, in the simplest case of fundamental vortex-nexus combinations, integral charge is found because the combination of fractional charge $1-1/N$ occurs together with fractional charge $1/N$, whose sum is 1.  Yet these two fractional charges are not otherwise correlated; they can and generally will be far apart, because entropy considerations cause vortices and nexus world lines to be very extended and convoluted; there is no long-range correlation between fractional topological charges except on a global scale.  

There are still fermionic zero modes associated with each integral combination of topological charge, but these are, in effect, not the usual zero modes around an instanton, which fall off in all four directions of Euclidean space-time.  Instead, because the zero modes must feel two separated fractional charges, they are essentially three-dimensional, extending without appreciable change along some sort of world line joining the two fractional charges.  This is essential \cite{ct,gt86} in order that chiral symmetry breakdown yield a bilinear fermion condensate instead of a $2N-f$-fold fermion condensate.  

So it appears that the combination of center vortices and their associated nexuses have the potential to solve several problems at once:  A $\theta$ period of $2\pi$ (equivalent to only integral charge); a topological susceptibility correctly scaling with $N$; and the problem of getting a bilinear CSB condensate.  

There remain numerous issues to be resolved.  First, the nexus-vortex picture given in section IV is greatly simplified, and it remains to be verified by real calculations or simulations.  (A realistic large-$N$ d=4 simulation of gauge theory is probably not practical, but there are possible models in d=2 which can be studied for relatively large values of $N$.)  

Second, for the picture we have given to work, it is necessary either that conventional integrally-charged instantons are rare (with relative probability $O(1/N^2$) or even disallowed, by virtue of entropy-favored transitions to separated fractional-charge states.  The entropy-favored decay of instantons into fractional charges can be seen for $N=2$, where explicit forms were given \cite{ct} for half-charges resembling instantons but joined by a sphaleronic world line.  Standard entropy arguments favor the extension and randomization of this world line, leading to loss of long-range correlation or interaction between the half-charged objects.   No such explicit forms are known for $N>2$, and they should be sought.  

Third, there is the question of d=3 gauge theories where there are somewhat similar phenomena, since vortices and nexuses exist there, and so does fractional topological charge.  The Chern-Simons term is expressible as a linking number \cite{c94,c96,c97}.  We have noted that d=3 gauge theories pose special problems not seen in d=4; for example, adding a Chern-Simons term changes the equations of motion and the behavior of vortices \cite{cor96}, giving them twist or writhe (equivalent to self-linking).

{\em Note Added:}  After this work was completed I found a preprint of
Engelhardt and Reinhardt \cite{er}, who discuss the $SU(2)$ case of vortices with nexuses in terms similar to mine.  They do not discuss the generalization to $SU(N)$ nor the Witten-Veneziano problem.

\acknowledgements  I thank S. Cherkis for valuable conversations.

\newpage

\newpage
  
\begin{figure}[ht] 
\centerline{\epsfxsize 1.1 truein \epsfbox{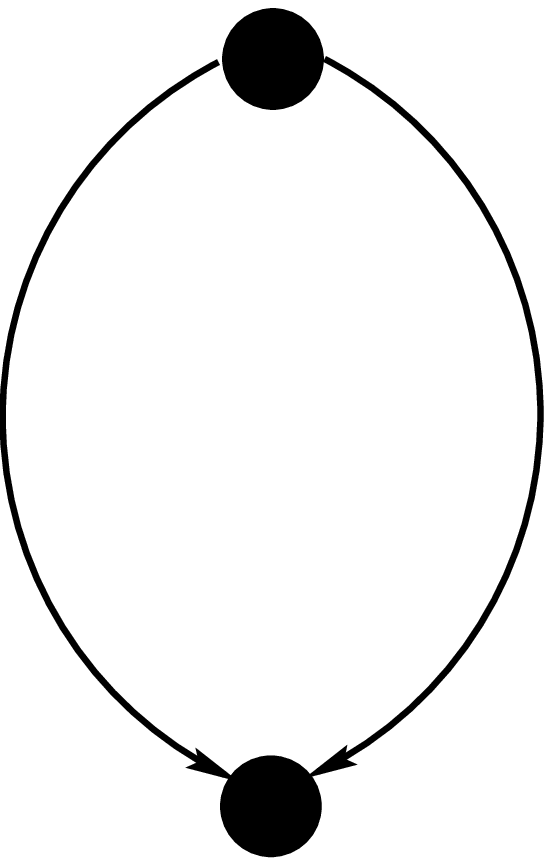}}   
\vskip.8 cm
\caption[]{
\label{fig1}
\small A sketch of an SU(2) nexus and anti-nexus (black circles) separating regions of center vortices with oppositely-directed field strengths.}
\end{figure}

\begin{figure}[ht]
\centerline{\epsfxsize 2 truein \epsfbox{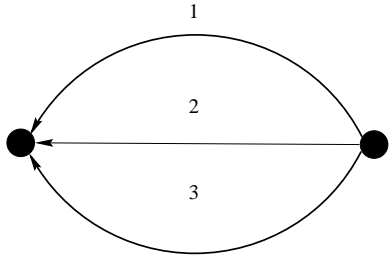}}
\vskip.8cm
\caption[]{
\label{fig2}
\small An SU(3) nexus and anti-nexus.}
\end{figure}

\begin{figure}[ht]
\centerline{\epsfxsize 3 truein \epsfbox{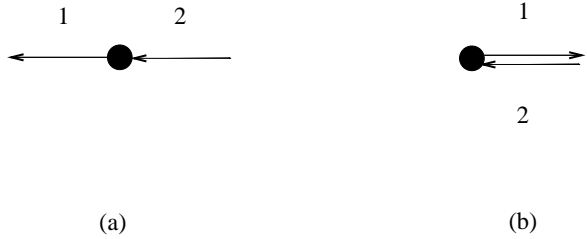}}
\vskip.8cm
\caption[]{
\label{fig3}
\small (a) A (-1,2) nexus of $SU(3)$ (isolated from its anti-nexus).  (b)  The (-1) line has been moved to merge with the (2) line, yielding a flux matrix $Q_2-Q_1$. }
\end{figure}

\begin{figure}[ht]
\centerline{\epsfxsize 3 truein \epsfbox{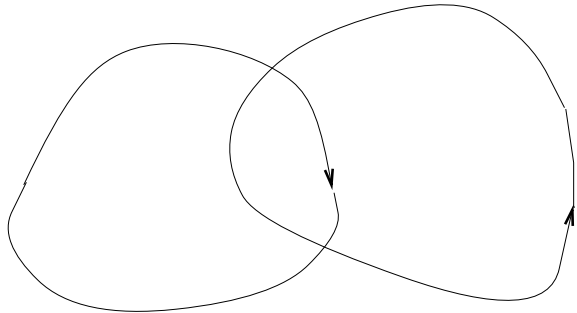}}
\vskip.8cm
\caption[]{
\label{fig4}
\small   In d=2, two closed lines intersect an even number of times, with opposite sign. }
\end{figure}

\begin{figure}[ht]
\centerline{\epsfxsize 3 truein \epsfbox{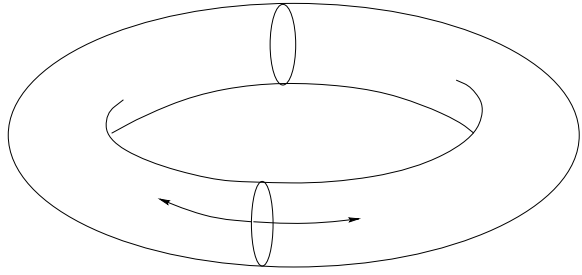}}
\vskip.8cm
\caption[]{
\label{fig5}
\small    Generation of a 2-surface in d=4 from the d=3 vortex-nexus combination of Fig.\ \ref{fig1}.  Note that the nexus world lines are now closed lines dividing the torus into regions of oppositely-directed field strength. }
\end{figure}

\begin{figure}[ht]
\centerline{\epsfxsize 3 truein \epsfbox{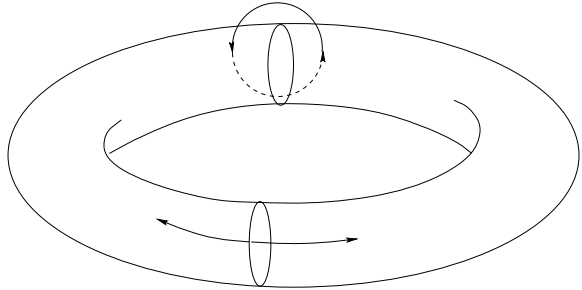}}
\vskip.8cm
\caption[]{
\label{fig6}
\small   Intersection (in d=4) of the vortex-nexus of Fig.\ \ref{fig2} with an ordinary vortex (slice shown as directed contour).  Linking of ordinary vortex with nexus world line produces topological charge. }
\end{figure}

\end{document}